\documentclass[aps,epsf,subfigure,twocolumn]{revtex4}

\usepackage{graphicx}% Include figure files
\usepackage{dcolumn}% Align table columns on decimal point
\usepackage{bm}% bold math
\usepackage{amsmath}
\usepackage{subfigure}
\usepackage{color,psfrag,epsfig}
\usepackage{multirow}
\usepackage{float}

\begin{document}

\newcommand{\up}{{\mid \uparrow \rangle}}
\newcommand{\down}{{\mid \downarrow \rangle}}

\title{Acoustic interactions between inversion symmetric and asymmetric two-level systems}

\author{A. Churkin$^{1,2}$}
\author{D. Barash$^2$}
\author{M. Schechter$^{1}$\footnote{smoshe@bgu.ac.il}}
\affiliation{$^1$Department of Physics, Ben Gurion University of
the Negev, Beer Sheva 84105, Israel}
\affiliation{$^2$ Department of Computer Science, Ben Gurion University of
the Negev, Beer Sheva 84105, Israel}

\date{today}

\begin{abstract}

Amorphous solids, as well as many disordered lattices, display remarkable universality in their low temperature acoustic properties. This universality is attributed to the attenuation of phonons by tunneling two-level systems (TLSs), facilitated by the interaction of the TLSs with the phonon field. TLS-phonon interaction also mediates effective TLS-TLS interactions, which dictates the existence of a glassy phase and its low energy properties. Here we consider KBr:CN, the archetypal disordered lattice showing universality. We calculate numerically, using conjugate gradients method, the effective TLS-TLS interactions for inversion symmetric (CN flips) and asymmetric (CN rotations) TLSs, in the absence and presence of disorder, in two and three dimensions. The observed dependence of the magnitude and spatial power law of the interaction on TLS symmetry, and its change with disorder, characterizes TLS-TLS interactions in disordered lattices in both extreme and moderate dilutions. Our results are in good agreement with the two-TLS model, recently introduced to explain long-standing questions regarding the quantitative universality of phonon attenuation and the energy scale of $\approx 1-3$ K below which universality is observed.

\end{abstract}

\maketitle

\section{Introduction}

The existence of two-level tunneling defects as a generic property in amorphous systems was postulated four decades ago\cite{AHV72,Phi72} in an attempt to explain the remarkable universality in the low energy characteristics of amorphous solids as were found earlier by Zeller and Pohl\cite{ZP71}. In what is now known as the Standard Tunneling Model (STM)\cite{AHV72,Phi72,Jac72} the interaction of the tunneling two-level systems (TLSs) with the phonon field is given by

\begin{equation}
H_{\rm TLS-ph} \;=\; \sum_j \sum_{\alpha,\beta} \;
\gamma_{j \rm s}^{\alpha \beta} S_j^z \; \frac{\partial u_j^{\alpha}}{\partial
{\bf r}_j^{\beta}} \, .
\label{impurityphonon}
\end{equation}

Here $S_j^z=\pm 1/2$ denote the two states of the TLS at site $j$, $u^{\alpha \beta} \equiv \partial u^{\alpha}/\partial{\bf r}^{\beta}$ is the phonon field, where $u$ is the phonon amplitude, $\alpha, \beta \equiv x,y,z$, and $\gamma$ is the TLS-phonon interaction parameter. The interaction of the phonon field with the tunneling amplitude of the TLS is small, and therefore neglected\cite{AHV72,Phi72}.

The TLS-phonon interaction as given in Eq.(\ref{impurityphonon}) gives rise to an effective TLS-TLS interaction mediated by the phonon field\cite{BH77,SS08}, which takes the form

\begin{equation}
H_{\rm TLS}^{\rm eff} = \sum_{ij} U_{ij} S_i^z
S_j^z
 \label{QIsing}
\end{equation}
where $U_{ij}$ is effectively random interaction with $1/r^3$ dependence for distance $r \gg a_0$. Here $a_0$ denotes the typical interatomic distance. These TLS-TLS interactions dictate the formation of a glassy phase in disordered lattices (see e.g. Ref.~\cite{HKL90}), and various low energy properties of the glassy phase in disordered lattices and amorphous solids. The STM, however, assumes that the TLSs are non-interacting, an assumption that is sufficient to explain many of the properties of amorphous solids in the universal regime, $T<T_U \approx 1-3$ K (notable exceptions are the phenomena of spectral diffusion\cite{BH77} and the dipole gap arising at very low temperatures, $T<0.1$K\cite{Bur95,RNO96,LNRO03}). Reconciliation of the assumption of non-interacting TLSs with the presence of TLS-phonon interaction is explained by the low density of TLSs, and consequent smallness of the typical interaction.

Various attempts were made to generalize the STM in an effort to explain quantitative universality, as well as phenomena such as the plateau in thermal conductivity between $\approx 2-10$ K shared between all amorphous solids and disordered lattices exhibiting the universal behavior. These attempts include specific theories for the disordered lattices\cite{KFAA78,SC85,GRS88,SK94}, theories of interacting TLSs\cite{YL88,Bur95,BNOK98,VL11}, and theories introducing soft potential modes\cite{KKI83,BGG+92,Par94,Kuhn03,PSG07}.

Recently a new model was suggested, which is based on the different properties of the interaction of inversion symmetric and inversion asymmetric TLSs with the phonon field\cite{SS09}. A TLS in which the two states are related to each other by inversion with respect to a midpoint, e.g. $180^{\circ}$ flip of CN impurity in a KBr lattice, does not interact with the phonon field, but only with the second derivative of the phonon amplitude\cite{SS08}

\begin{equation}
H^{\rm sym}_{\rm TLS-ph} = \sum_{j} \sum_{\alpha,\beta,\delta} \zeta_{\alpha
\beta \delta}({\bf r}_j) \frac{\partial^2 u_j^{\alpha}}{\partial
{\bf r}_j^{\beta}
\partial {\bf r}_j^{\delta}} \tau_j^z ,
 \label{H2int}
\end{equation}
where we denote by $\tau$ the symmetric TLSs, to differ from the asymmetric TLSs denoted by $S$. This form of the interaction leads, in three dimensions, to a long distance dependence of $1/r^4$ and $1/r^5$ for the acoustic mediated $S$ TLS-$\tau$ TLS (S-$\tau$) effective interactions and $\tau$ TLS -$\tau$ TLS ($\tau$-$\tau$) effective interactions respectively\cite{SS08}. In two dimensions similar analysis to the one carried in Ref. \cite{SS08} leads to $1/r^2, 1/r^3, 1/r^4$ spatial dependence of the S TLS -S TLS (S-S), S - $\tau$, and $\tau$ - $\tau$ interactions respectively.

Where the above is correct for $2$ impurities in an otherwise pure lattice, the presence of disorder introduces deviations from inversion symmetry, and a finite, albeit small interaction between a $\tau$-TLS and the phonon field. In Ref.\cite{SS09} it was argued that for systems with strong disorder the TLS-phonon interaction can be written as

\begin{equation}
H^{\rm tot}_{\rm TLS-ph} \;=\; \sum_{j,\alpha,\beta} \;
\gamma_{\rm s}^{\alpha \beta} ({\bf r}_j) S_j^z \; u_j^{\alpha \beta} + \sum_{j',\alpha,\beta} \;
\gamma_{\rm w}^{\alpha \beta} ({\bf r}_{j'}) \tau_{j'}^z \; u_{j'}^{\alpha \beta} .
\label{Stauphonon}
\end{equation}
The small dimensionless parameter of the theory\cite{SS09} is defined by $g \equiv \gamma_{\rm w}/\gamma_{\rm s}$. This parameter is proportional to the deviations from inversion symmetry, and therefore to the ratio between the strain and interatomic lattice spacing, i.e. in strongly disordered systems $g \approx 0.01-0.03$. The resulting effective elastic TLS-TLS interaction takes the form\cite{SS09}

\begin{equation}
H_{\rm S\tau}^{\rm eff} = \sum_{ij} U_{ij}^{SS} S_i^z
S_j^z + \sum_{ij} U_{ij}^{S \tau} S_i^z
\tau_j^z + \sum_{ij} U_{ij}^{\tau \tau} \tau_i^z
\tau_j^z
 \label{effectiveStau}
\end{equation}
where in three dimensions (two dimensions) all interactions decay as $1/r^3$ ($1/r^2$) at distances $r \gg a_0$, and their typical values at near neighbor distance are related by

\begin{equation}
U_0^{\tau \tau} \approx g U_0^{S \tau} \approx g^2 U_0^{S S} .
\label{gratios}
\end{equation}
It is further argued\cite{SS09} that the above long distance power law dependence of the interaction persists down to a short distance cutoff, which is not much larger than the interatomic spacing, for lack of another relevant length scale in the system.
The form of the Hamiltonian ({\ref{effectiveStau}), ({\ref{gratios}), i.e. the strength and spatial dependence of the three TLS-TLS interactions, and its subsequent analysis,
allow for an explanation of some long standing questions related to the low temperature universality in glasses\cite{SS09}. The universality and smallness of phonon attenuation\cite{HR86,PLT02} are shown to be a consequence of the generic characteristics of the $\tau$ TLSs, and the energy scale of $\approx 1-3$ K dictating the universal regime
is attributed to the gapping of the S-TLSs below the energy of $U_0^{S \tau} \approx g U_0^{S S} \approx g T_G \approx 1-3$ K, where $T_G$ is the glass temperature\cite{note_G}. The enhancement of the single particle S-TLS DOS at higher temperatures results in their domination of acoustic attenuation, and the limiting of the universal characteristics of amorphous and disordered solids to $T < T_U$.

The two-TLS model further derives, and quantifies, some of the central assumptions of the STM. The DOS of the relevant TLSs at low temperatures, the $\tau$-TLSs, is found to be finite at very low energies and rather homogeneous for energies smaller than $3$K\cite{SS09,CGBS13,CBS14}. The assumption of noninteracting TLSs is supported by the fact that at low temperatures the S-TLSs are frozen, and the typical $\tau-\tau$ interaction $U_0^{\tau \tau} \approx 0.1$K is smaller by a factor of $g$ in comparison to the typical $\tau$-TLS energy, the latter dictating the typical energy scale of the universal phenomena. Thus, within the two-TLS model TLS-TLS interactions can be neglected except at very low temperatures. However, this possibility to neglect the interactions does not depend on the small concentration of TLSs, but on the form of the Hamiltonian ({\ref{effectiveStau}), ({\ref{gratios}). This is in line with the existence of universal phenomena in disordered lattices such as KBr:CN, where there are strong evidences to the notion that each CN impurity constitutes a two-level system. Thus, the TLS-DOS in disordered lattices is not small, but the relevant TLSs are $\tau$-TLSs, between which the interaction is small and can therefore be neglected, and the experimentally observed smallness of the TLSs DOS is related to the fact that only TLSs with appreciable tunneling amplitude can be detected. Note that in KBr:CN the TLS is formed by each CN low energy state and the state related to it by a CN flip\cite{SC85,SK94,SS09,GS11}, and is thus a $\tau$-TLS, where all other single particle states of a given CN impurity are much higher in energy\cite{CBS14}, as they constitute an S excitation with respect to the CN low energy state\cite{SS09}.

The two-TLS model, and specifically the Hamiltonian ({\ref{effectiveStau}), was derived microscopically for strongly disordered lattices. However, the strong evidence for the equivalence of the phenomenon of the low temperature universality in disordered lattices and amorphous solids\cite{LVP+98,PLC99,TRV02,PLT02}, 
%(as well as other disordered systems, e.g. disordered polymers, metallic glasses, quasicrystals) 
suggests that the mechanism leading to universality in disordered lattices and amorphous solids is the same. Thus, validation of the two-TLS model may prove useful both for the resolution of the long standing problem of universality in disordered and amorphous solids, and for the advance of our understanding of the microscopic structure of amorphous solids. The two crucial points of the two-TLS model are the form of the Hamiltonian ({\ref{effectiveStau}), ({\ref{gratios}), and the resulting structure of the DOS of the $\tau$ and $S$ TLSs. With regard to the latter, the results in Ref.\cite{SS09} are supported by a comprehensive numerical calculation given in Refs.\cite{CGBS13,CBS14}. With regard to the former, a significant step was made in Ref.\cite{GS11}, where $\gamma_{\rm w}, \gamma_{\rm s}$ were explicitly calculated for KBr:CN. The obtained ratio of $\gamma_{\rm w}/\gamma_{\rm s} \approx 0.02$ and value $\gamma_{\rm w} \approx 0.1$eV are both in agreement with theory\cite{SS09} and with the experimental value for the coupling constant for the relevant TLSs at low temperatures,\cite{BDL+85,YKMP86}. These results support both the two-TLS model and the notion that it is indeed CN flips that constitute the TLSs for KBr:CN. However, the ab-initio and DFT calculations\cite{GS11} required the use of very small samples, not allowing the study of the effective TLS-TLS interactions and their distance dependence.

In this paper we use the conjugate gradients method to study the magnitude and spatial dependence of the impurity-impurity interactions in KBr:CN.
Calculations are performed for S-S, S-$\tau$, and $\tau$-$\tau$ interactions in two and three dimensions,
as function of distance between the TLSs.
The obtained distinct magnitude and power law spatial dependence of the three interactions, as well as disorder induced change in the power law of the S-$\tau$ interactions, which occurs at distances not much larger than the inter-atomic spacing, are in agreement with the two-TLS model.
In Sec. \ref{sec:model} we present the calculation setup. Our results are presented in Sec. \ref{sec:results}, and conclusions in Sec. \ref{sec:conclusions}.

\section{The Model}
\label{sec:model}

All our calculations are carried for CN impurities in KBr lattice.
We arrange initially a 3D grid of volume $N \times N \times N$  ($N \times N$ in 2D, $N$ is even)  of $K$\textsuperscript{+} and $Br$\textsuperscript{-} ions, having distance $3.1974\AA$  ($3.2735\AA$ in 2D)  between the ions. These distance values are calculated by the energy minimization procedure of pure KBr grid.
Then we replace some of the $Br$\textsuperscript{-} ions with $CN$\textsuperscript{-} ions. The $K$\textsuperscript{+} and $Br$\textsuperscript{-} ions are assumed to carry +1 and -1 charges respectively, while the charge of the $CN$\textsuperscript{-} ion is represented by fractional charges q$_{C}=-1.28$ and q$_{N}=-1.37$ placed on the carbon and nitrogen atoms, and the additional charge q$_{center}=+1.65$ is placed at the center of mass \cite{KM83}. The C-N bond length is fixed at $1.17\AA$, while the distances of the carbon and nitrogen atoms from the center of mass are fixed at $0.63\AA$ and $0.54\AA$ respectively \cite{KM83}. Interatomic potential is calculated by the formula:

\begin{equation}
V_{\alpha \beta}(R) = A_{\alpha \beta}\exp(-a_{\alpha \beta}r) + {B_{\alpha \beta} \over r^6} + K{q_{\alpha}q_{\beta} \over r}.
\label{potential}
\end{equation}
The interatomic potential parameters A$_{\alpha \beta}$, $a_{\alpha \beta}$ and B$_{\alpha \beta}$ are taken from Ref. \cite{BKMO82} and shown in Table~\ref{tab:params}, and $K=1389.35 \AA$kJ/mol.

\begin{table}[t]
\def\arraystretch{1.5}
\begin{tabular}{|c|c|c|c|}
	\hline
$_{\alpha \alpha}$ &  $A_{\alpha \alpha}$ (kJ/mol)  &  $a_{\alpha \alpha}$ (1/$\AA$) & $B_{\alpha \alpha}$ ($\AA^6$kJ/mol)   \\
	\hline
KK & 158100 & 2.985 & -1464 \\
CC & 259000 & 3.600 & -2110 \\
NN & 205020 & 3.600 & -1803 \\
BrBr & 429600 & 2.985 & -12410 \\
	\hline
\end{tabular}
\caption {Interatomic potential parameters (taken from Ref. \cite{BKMO82}). Cross-interaction parameters were calculated by $A_{\alpha \beta}=(A_{\alpha \alpha}A_{\beta \beta})^{1/2}$, $a_{\alpha \beta} = (a_{\alpha \alpha} + a_{\beta \beta})/2$, $B_{\alpha \beta}=-(B_{\alpha \alpha}B_{\beta \beta})^{1/2}$. }
\label{tab:params}
\end{table}

Non-linear conjugate gradients method (Fletcher–Reeves) is used to find the closest local minimum of the system potential energy U. Periodic boundary conditions are used to simulate the infinite crystal.

In order to study the spatial dependence of the CN-CN interactions in the KBr:CN crystal we calculate U
for KBr:CN grids containing first only two CN ions. The CNs are placed at some distance $R$ from each other in some direction [$x,y,z$]. For example, if the first CN ion is placed at coordinate [$0,0,0$] in the grid (3D array), $R=3$ and the direction is [2,2,2], then the second CN is placed at [$R \cdot x,R \cdot y,R \cdot z$] = [$6,6,6$]. For a given distance $R$
and direction [$x,y,z$] we calculate the energies for several different orientations of CN ions in the grid and we use these energies to evaluate the S-S, S-$\tau$ and $\tau$-$\tau$ interactions, as explained in detail below. Because the non-linear conjugate gradients method is very time consuming for large samples, we use both two dimensional and three dimensional grids. The three dimensional grids relate to the relevant experimental systems, whereas the two dimensional grids allow us to study spatial dependence of CN-CN interactions at longer distances, crucial especially for the observation of the effects of disorder, see below.

\begin{figure}[tb!]
\centering
\includegraphics[width=9cm]{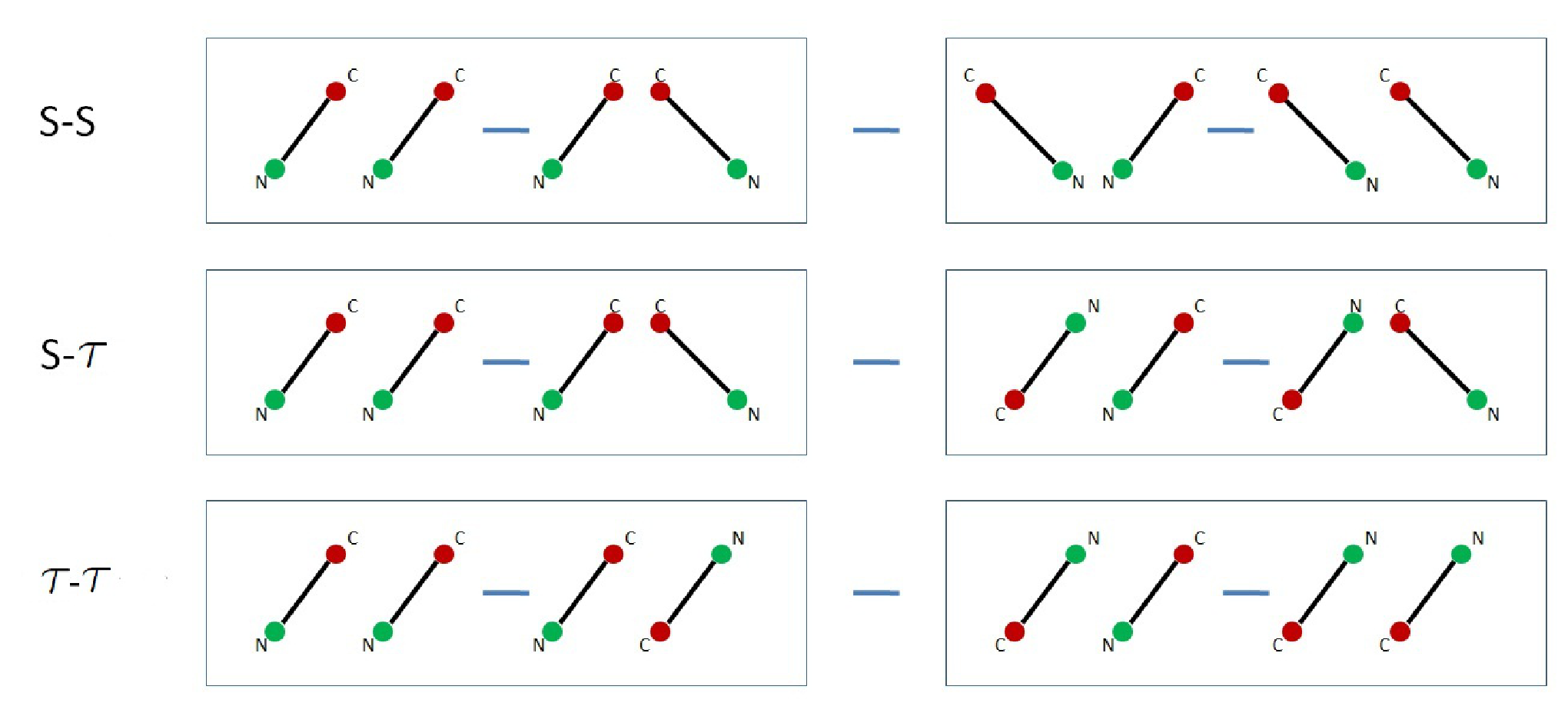}
\caption{ S-S, S-$\tau$ and $\tau$-$\tau$ interactions calculation in 2D grid with two CN ions placed at some distance r from each other. Each pair of CN ions in the figure represents the energy U$_{r}$ reached by local minimization after orienting the CN ions in the direction shown in the figure. The interaction energy U for each of the three types of interactions is calculated by U=(U$_{r}(1)$-U$_{r}(2)$)-(U$_{r}(3)$-U$_{r}(4)$).}
\label{fig:inter}
\end{figure}

We start by finding the low energy states for a single CN impurity in an otherwise pure KBr:CN lattice.
In two dimensions these low energy states are four-fold degenerate, along the square diagonals. In three dimensions these states are eight-fold degenerate along the in-space diagonals.
We emphasize that although for a single CN impurity the ground state degeneracy is lifted by tunneling, for strong disordered systems (CN concentration $0.2<x<0.7$) the CN-CN interactions are much larger than the tunneling amplitude, and therefore it is the interaction with neighboring CN impurities that breaks the single impurity ground state degeneracy. The treatment of tunneling can then be deferred until the bias energies are determined\cite{SS09}.
Since in this paper we are concerned only with the energy bias of the different orientational states resulting from interactions with other CN impurities, tunneling is neglected altogether. This is also in line with our chosen calculation method, which does not allow tunneling over large barriers\cite{note}.

Let us now describe the calculation of the interaction, starting with two dimensions, see also Fig.~\ref{fig:inter}. Both CN ions are placed initially at relative distance $r$, and pointing in directional angles
${\phi_1,_2 = \pi/4}$ (angles are calculated between the direction of the N-C vector and the x-axis). The whole system is then relaxed. The CN-CN interaction results in a small shift in the CN positions ( $\ll a_0$) and angles ( $\ll \pi$). The resulting energy of the system, U$_{r}(1)$ and positions of the CN molecules are recorded.
We then follow the same procedure to calculate U$_{r}(2)$ with the first CN ion having ${\phi_1 \approx \pi/4}$ and the second ion is rotated with ${\phi_2 \approx 3\pi/4}$, U$_{r}(3)$ with the first CN ion rotated having ${\phi_1 \approx 3\pi/4}$ and the second ion with ${\phi_2 \approx \pi/4}$ and U$_{r}(4)$ having both CN ions rotated with ${\phi_1,_2 \approx 3\pi/4}$. Finally, the S-S interaction energy is calculated by: U$_{r}$(S-S) $=$ (U$_{r}(1)$ - U$_{r}(2)$) - (U$_{r}(3)$ - U$_{r}(4)$). S-$\tau$ and $\tau$-$\tau$ interactions energies are calculated in a similar way. The difference in S-$\tau$ and $\tau$-$\tau$ interactions calculation is that U$_{r}(3)$ and U$_{r}(4)$ are calculated while the first CN ion instead of being rotated is flipped, i.e. ${\phi_1 \approx 5\pi/4}$. Additionally, for the $\tau$-$\tau$ calculation the second CN in U$_{r}(2)$ and U$_{r}(4)$ is flipped: ${\phi_2 \approx 5\pi/4}$, see Fig.~\ref{fig:inter}.

\newcommand{\specialcell}[2][c]{%
  \begin{tabular}[#1]{@{}c@{}}#2\end{tabular}}

\begin{table}[t]
\def\arraystretch{1.5}
\begin{tabular}{|c|c|c|c|c|c|}
	\hline
Interaction & Angle &  U$_{r}$1 & U$_{r}$2 & U$_{r}$3 & U$_{r}$4  \\
	\hline
\multirow{4}{*}{S-S} & $\phi_{(CN1)}$ & $\pi/4$ & $3\pi/4$ & $\pi/4$ & $3\pi/4$ \\
  & $\theta_{(CN1)}$ & 0.955 & 0.955 & 0.955 & 0.955 \\
  & $\phi_{(CN2)}$ & $\pi/4$ & $\pi/4$ & $3\pi/4$ & $3\pi/4$ \\
  & $\theta_{(CN2)}$ & 0.955 & 0.955 & 0.955 & 0.955 \\
    \hline
\multirow{4}{*}{S-$\tau$} & $\phi_{(CN1)}$ & $\pi/4$ & $3\pi/4$ & $\pi/4$ & $3\pi/4$ \\
  & $\theta_{(CN1)}$ & 0.955 & 0.955 & 0.955 & 0.955 \\
  & $\phi_{(CN2)}$ & $\pi/4$ & $\pi/4$ & $5\pi/4$ & $5\pi/4$ \\
  & $\theta_{(CN2)}$ & 0.955 & 0.955 & 2.186 & 2.186 \\
	\hline
\end{tabular}
\caption {Approximate values of $\phi$ and $\theta$ angles of two CN ions in the three dimensional grids, used in the calculation of U$_{r}(i)$ for S-S and S-$\tau$ interactions.}
\label{tab:ang3D}
\end{table}

In three dimensions the calculations are carried in a similar way.
Since the low energy states of a single CN impurity are eight-fold degenerate,
each CN impurity, whereas it can assume only a single flip, can rotate in six different directions. We thus specify for each calculation which of the allowed rotations is performed, see Table~\ref{tab:ang3D}.

As mentioned above, the spatial dependence of the S-$\tau$ and $\tau$-$\tau$ elastic interactions is modified in the presence of disorder. We therefore repeat the above calculations for CN-CN interactions in the presence of a third impurity. As the purpose of the added impurity is to introduce strain disorder, it is placed in a given position and orientation, and is then relaxed with the whole system for each configuration of the two ``original'' CNs between which the interaction is calculated.

All our calculations are done for the total interaction energy, which includes the acoustic mediated interaction and the electric dipolar interaction. With regard to the S-S interaction and S-$\tau$ interaction the electric dipolar interaction is significantly subdominant, and thus our results can be directly compared to theory done for the acoustic mediated interaction. With respect to the $\tau$-$\tau$ interaction, the electric dipolar interaction and the acoustic mediated interaction are of similar order. Thus, our results for the $\tau$-$\tau$ interaction reflect well their approximate magnitude, but should be considered with caution with respect to their functional dependence.

\section{Results}
\label{sec:results}

The S-S, S-$\tau$, and $\tau$-$\tau$ interaction energies [U(S-S), U(S-$\tau$) and U($\tau$-$\tau$)] as function of distance for two CN impurities in an otherwise pure two dimensional lattice are plotted in Fig. \ref{fig:2D}. The CN impurities are placed in directions [$2,0$] and [$3,1$], and calculations are done for
grids of size $50 \times 50$ ions ($25 \times 25$ unit cells),
up to maximal distance of $11$ unit cells, as at larger distances boundary condition effects become significant.

\begin{figure}[tb!]
\centering
\includegraphics[width=9cm]{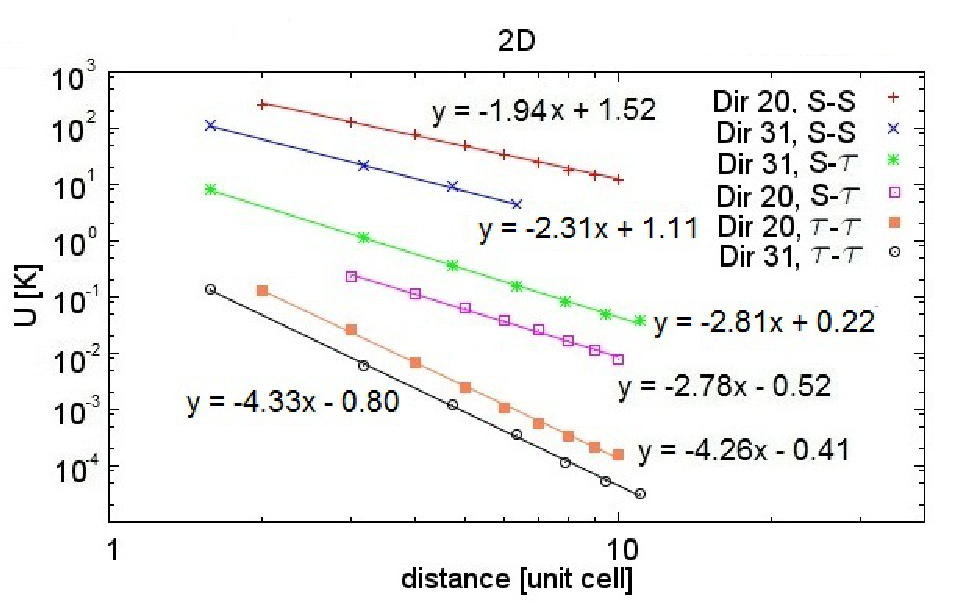}
\caption{ Spatial dependence of the elastic S-S, S-$\tau$ and $\tau$-$\tau$ interactions of 2 CNs placed in directions [2,0] and [3,1] in 2D grid of size $50 \times 50$ ions ($25 \times 25$ unit cells). Here and in subsequent figures energies are given in Kelvin units, $1 {\rm (KJ/mol}) / k_{\rm B} = 120.274$ K. Solid line are best linear fits for the Log-Log plots, where slopes denote the power law distance dependence of the interaction. }
\label{fig:2D}
\end{figure}

\begin{figure}[tb!]
\centering
\includegraphics[width=9cm]{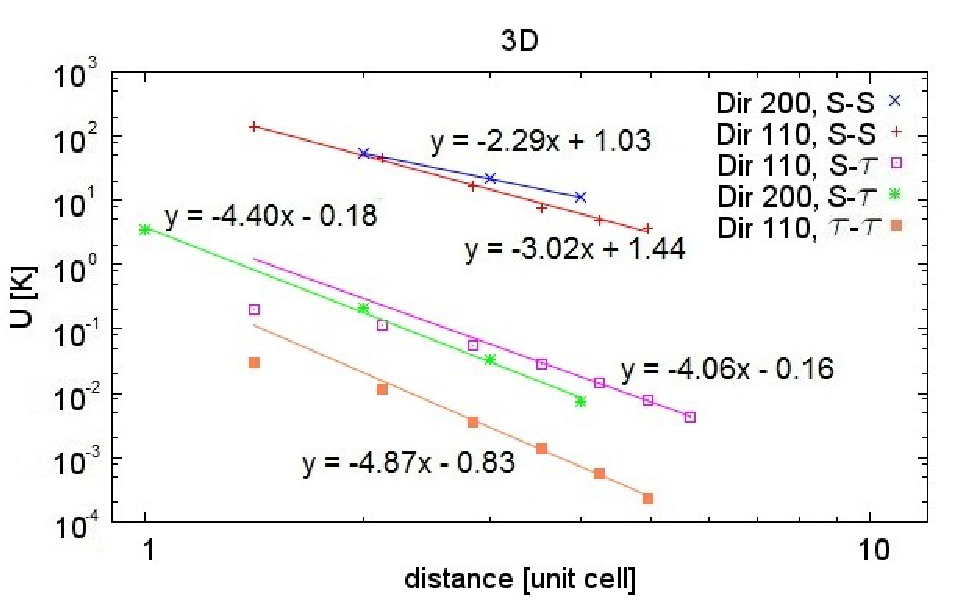}
\caption{ Spatial dependence of the elastic S-S, S-$\tau$ and $\tau$-$\tau$ interactions of 2 CNs placed in directions [2,0,0] and [1,1,0] in 3D grids of size $20 \times 20 \times 20$ ions ($10 \times 10 \times 10$ unit cells). Asymptotic fits for the data at the largest available distances are given.}
\label{fig:3D}
\end{figure}

Since our available distance range is limited, lattice discreteness is expected to play a role. Indeed, the exact values of the magnitude of the interactions at short distance and their distance dependence depends on details such as the direction between the two CN impurities. Still, our numerical results strongly support both the typical magnitude of the various elastic interactions (S-S, S-$\tau$, $\tau$-$\tau$) at short distance, and their spatial dependence, as discussed above.

The magnitude of the interactions we find numerically is in agreement with experiments, and with their interpretation by the two-TLS model. The elastic S-S interaction at short distance is of the order of $100$K, which is the same order of the glass temperature in KBr:CN. The elastic S-$\tau$ interaction is of the order of $3$K, which is $g$ times smaller, and is characteristic of the temperature below which universality is observed\cite{SS09}. The elastic $\tau$-$\tau$ interaction is of the order of $0.1$K (another factor of $g$ smaller), in agreement with the theory of the two-TLS model, and with low energy experiments reporting interaction phenomena at this energy scale\cite{RNO96,LNRO03}. The spatial dependence is also in good agreement with theory\cite{SS08}, as the S-S, S-$\tau$, and $\tau$-$\tau$ elastic interactions behave as $1/r^{\alpha}$, with $\alpha \approx 2,3,4$ respectively. We note that this power law dependence of the various interactions persists down to very short distances, not much larger than the interatomic spacing.

\begin{figure}[tb!]
\centering
\includegraphics[width=9cm]{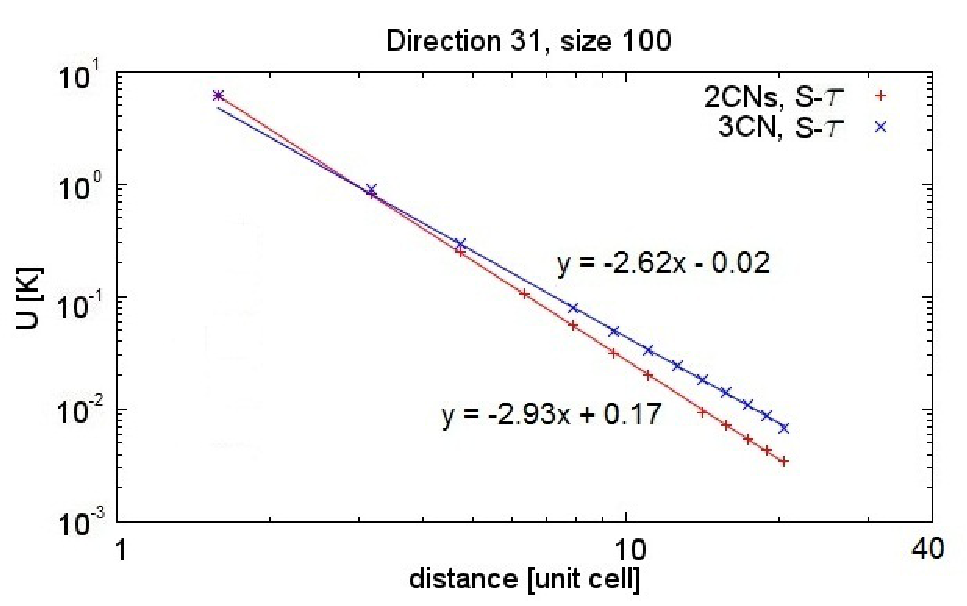}
\caption{ Spatial dependence of the elastic S-$\tau$ interactions of two CNs placed in the direction [3,1] in 2D grid of size $100 \times 100$ ions ($50 \times 50$ unit cells) in the absence and in the presence of a third impurity. The third impurity is positioned as near neighbor in direction [1,1] with respect to the ``$\tau$" impurity (the impurity for which we study its flip), and with orientation $\phi=3\pi/4$. Data for 2 CNs corresponds to S-$\tau$ interaction in a pure KBr lattice. Data for 3 CNs corresponds to S-$\tau$ interaction (calculated for impurities 1 and 2) in a disordered lattice (disorder is introduced by impurity 3).
}
\label{fig:31_100}
\end{figure}

Similar calculation
were performed for three dimensional samples of size $20 \times 20 \times 20$ ions ($10 \times 10 \times 10$ unit cells), for CN impurities in relative directions [1,1,0] and [2,0,0], see Fig. \ref{fig:3D}.
Our calculations in three dimensions are limited to distances of six unit cells. Still, good agreement with theory is obtained. The typical magnitudes of the S-S, S-$\tau$, and $\tau$-$\tau$ interactions at short distance are similar to those obtained in two dimensions, but the powers of the functional dependence of the interactions is raised by one because of the extra dimension.

A crucial component of the two-TLS model is the reduction in power law for the $S-\tau$ interaction upon the inclusion of disorder. To study this change in functional dependence we repeat our calculations for the interaction of CN impurities at relative direction [3,1] with the inclusion of a third impurity. The third impurity is placed as near neighbor in direction [1,1] with respect to the ``$\tau$"-TLS, the one that performs the flip, as it is the deviation from local inversion symmetry induced by strain disorder at the position of the $\tau$-TLS that causes the change in functional dependence of the interaction. As the deviations induced by the third impurity to the $S-\tau$ interaction are insignificant at very short distances, we have performed these calculations on an extended lattice in two dimensions, of size $100 \times 100$ ions ($50 \times 50$ unit cells). Enhancement of the interaction resulting from the inclusion of disorder is observed already at the rather modest distances studied, with a clear shift in the power law of the interaction, see Fig.\ref{fig:31_100}. Our results here also support the theoretical prediction that disorder does not affect the typical magnitude of the elastic S-$\tau$ interaction at short distance.

\section{Conclusions}
\label{sec:conclusions}

Using conjugate gradients method we have calculated the elastic interactions between rotations and flips of CN impurities in a KBr lattice. CN flips constitute two states related to each other by inversion symmetry, defined as $\tau$ TLSs. CN rotations constitute asymmetric (S) TLSs. For two CN impurities in an otherwise pure lattice we find that the elastic S-S, S-$\tau$, and $\tau$-$\tau$ interactions differ both in their magnitude at short distances, and in their spatial power law dependence. Introducing disorder does not change the relative magnitude of the interactions at short distances, but affects the spatial dependence of interactions involving a $\tau$-TLS, as inversion symmetry is broken. Our results are in agreement with theoretical predictions both for the magnitude and the power law dependence of the interactions. Further, we find that the power law dependence predicted for large distances persists down to very short distances, not much larger than the interatomic distance. Our results can serve as basis for the calculation of interaction dominated properties in disordered lattices in both very low concentrations, where disorder can be neglected, and in moderate concentrations, where disorder plays a significant role. 

In addition, the magnitude and spatial dependence of the elastic S-$\tau$ interactions is a crucial element in the two-TLS model\cite{SS09}. Whereas the two-TLS model is microscopically derived for the disordered lattices, it explains the smallness and universality of the tunneling strength and the energy scale $\approx 1$K as they appear equivalently in disordered lattices and amorphous solids. Validation of the applicability of the two-TLS model to amorphous solids is thus of much interest, and requires the reconciliation of its arguments based on small deviations from local inversion symmetry with the non existence of long range order in amorphous solids and with the indications of tunneling states being composed of $\approx 10-50$ atoms in typical amorphous solids\cite{BZN+88,BGGS91,LS91,HS96,PSG07}. Such research is now in progress, and if successful,
%for the low temperature universality in disordered and amorphous solids.
% 
%Thus, 
our results here will contribute to the 
%validation of the above model, and as such may 
advance of our understanding not only of the interactions between tunneling states in disordered solids, but also of the microscopic structure of amorphous solids and of the low temperature properties of glasses.

\section{Acknowledgments}

We thank A. Gaita-Ari\~no, N. Gr{\o}nbech-Jensen, and J. Rottler for very useful discussions. This research was supported by the Israel Science Foundation (Grant No. 982/10).


\begin{thebibliography}{10}
\bibitem{note_G}
This argument produces the correct energy scale for universality, and is explicitly verified for KBr:CN\cite{CBS14}. However, the similarity of the universal temperature between systems with rather different glass transition temperature is stronger than this argument can account for, and requires further consideration.
\end{thebibliography}
\end{document}